\documentclass{pasj00}
\begin{document}
\SetRunningHead{K. Koyama}{A Time Variable X-Ray Echo}
\Received{2007/06/20}
\Accepted{2007/08/24}
\title{A Time-Variable X-Ray Echo: Indications of a Past Flare of the Galactic-Center Black Hole }
\author{Katsuji \textsc{Koyama}, Tatsuya \textsc{Inui}, Hironori \textsc{Matsumoto}, and Takeshi Go \textsc{Tsuru} }
\affil{Department of Physics, Graduate school of Science, Kyoto University, Sakyo-ku, Kyoto 606-8502}
\email{koyama@cr.scphys.kyoto-u.ac.jp}

\maketitle

\begin{abstract}
A time-variability study of the neutral iron line flux at 6.40~keV in the Sgr B2 region from data of 
Suzaku and  Chandra is presented. 
The highly ionized iron line at 6.68~keV is due to Galactic Center Diffuse X-rays (GCDX), and is thus  time invariable.
By comparing the 6.68 keV and 6.40 keV line fluxes, we found
that the 6.40 keV flux from the Sgr~B2 complex region is time variable; 
particularly the giant molecular cloud 
M~0.66$-$0.02, known as ``Sgr B2 cloud'' is highly variable.
The variability of the 6.40~keV line in intensity and spatial distribution strongly supports the scenario that 
the molecular clouds in the Sgr B2 region are X-ray Reflection Nebulae irradiated by the Galactic Center (GC) 
black hole Sgr A$^*$. 
\end{abstract}

\section{Introduction}

Accurate astrometric and spectroscopic measurements of the Galactic Center (GC) stars revealed that the GC exhibits a $4 \times10^6$ solar-mass black hole at the gravitational center Sgr A$^*$
(Sch\"{o}del et al. 2002; Ghez et al. 2003).
  Massive black holes  are usually violently bright at all  wavelengths, particularly 
in X-rays, and hence are called Active Galactic Nuclei (AGN).  Sgr A$^*$, however, is very quiescent 
with an X-ray luminosity of only 10$^{33-34}$ ergs s$^{-1}$ (Baganoff et al. 2001), 
many orders of magnitude dimmer
than that of a canonical AGN.  A question is whether or not Sgr A$^*$ has always been quiescent.

The radio complex Sgr~B2 comprises many H~\emissiontype{II} regions and molecular clouds.
\citet{Koyama1996} and Murakami et al.(2000) found that molecular clouds emit peculiar X-rays with a strong 6.40~keV line 
(equivalent width of 1--2~keV) and  deep iron absorption edge at 7.1~keV (equivalent $N_{\rm H}$ 
of $\sim10^{24}$~H~cm$^{-2}$).
A detailed morphology of the Sgr~B2 cloud has been studied with 
Chandra \citep{Murakami2001}, and other 6.40~keV structures (e.g., M~0.74$-$0.09) have been found 
with Suzaku (Koyama et al. 2007b), both of which indicate that the Sgr~B2 cloud and its complex are X-ray 
reflection nebulae (XRNe) irradiated by the GC source Sgr~A$^*$; Sgr~A$^*$ was X-ray bright about 300 years ago, the light-travel time between  Sgr~B2 and Sgr~A$^{*}$.

A counterargument of the XRN scenario is that the origin of the 6.40~keV line is due to 
charged particles (e.g. Valinia et al. 2000; Yusef-Zadeh et al. 2007). 
A direct test to distinguish these two possibilities is to detect any time variability of the 6.40 keV line flux.
Muno et al.(2007) reported the Chandra discovery of
a 2--3 year's variability of the 6.40 keV line within 20 pc of Sgr A$^*$.
This observation suggests that Sgr A$^*$ had an  X-ray
luminosity of $\sim10^{38}$ ergs~s$^{-1}$ around 60 years ago.
This paper reports on a longer-term (5-years) time variability from an extended region of the Sgr~B2 cloud. The errors quoted in this paper are at the 90\% confidence level, unless otherwise mentioned.

\section{Observation and Data Reduction}

\subsection{The Chandra Observation}

Sgr B2 was observed using the Advanced CCD Imaging Spectrometer (ACIS-I) aboard the Chandra observatory \citep{Weisskopf2002} on 2000 March 
29--30. The total exposure time, after data screening was 100 ksec.
 We used the event file provided by standard pipeline processing.  Only the ASCA grades 0, 2, 3, 4 and 6 events were used in the analysis.
The details of the data reduction are the same as those of Murakami et al.(2001). 
ACIS-I consists of four CCDs, each  covering  an 8.3-arcmin square on the sky, while the pixel size is 0.5 arcsec. The on-axis spatial resolution is 0.5-arcsec full 
width at half-maximum (FWHM). Images, spectra, ancillary response  file (ARF) and response matrices have been created using CIAO~v3.4 software. The absolute positional accuracy was 0.6~arcsec \citep{Weisskopf2003}.

\subsection{The Suzaku Observation}

The Sgr B2 region was observed with  XIS on 2005 October 10--12. XIS consists of four sets of X-ray CCD camera systems (XIS0, 1, 2, 
and 3) placed on the focal planes of four X-Ray Telescopes (XRT) aboard the Suzaku satellite. XIS0,
2 and 3 have front-illuminated (FI) CCDs, while XIS1 has a 
back-illuminated (BI) CCD. Detailed descriptions of the Suzaku satellite, XRT 
and XIS can be found in
\citet{Mitsuda2007}, \citet{Serlemitsos2007} and Koyama et al.(2007a), respectively.
The XIS observation was made in the normal mode. The effective exposure time after removing the epoch of the low-Earth elevation angle (ELV 
$\le$5$^\circ$) and
the South Atlantic Anomaly was about 89~ksec. We analyzed the data using the software package HEASoft 6.2. We fine-tuned the XIS astrometry using the Chandra point sources in the field of view. The details of the data reduction are the same as  that of Koyama et al.(2007b). 

\section{Analysis and Results}

\subsection{The Sgr B2 Complex}
We selected the Sgr B2 region, which two observations commonly covered. 
The region is shown by the solid line in figure 1 (upper panel). 
\begin{figure}[!ht] 
\begin{center}
\vspace*{5mm}
\FigureFile(65mm,70mm){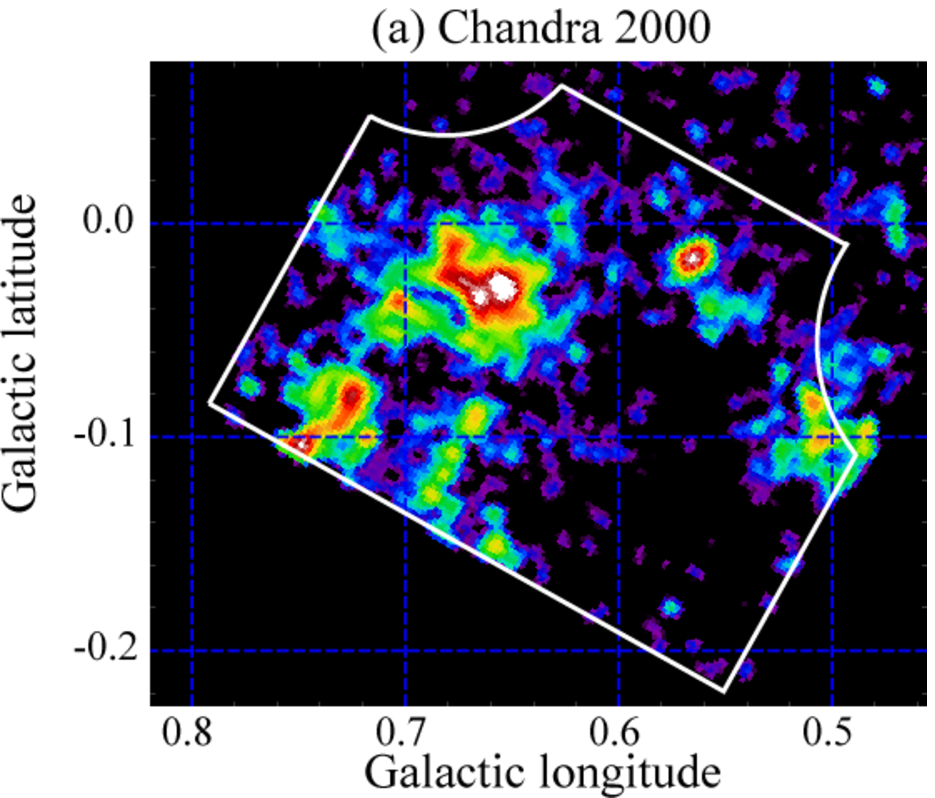}
\FigureFile(65mm,70mm){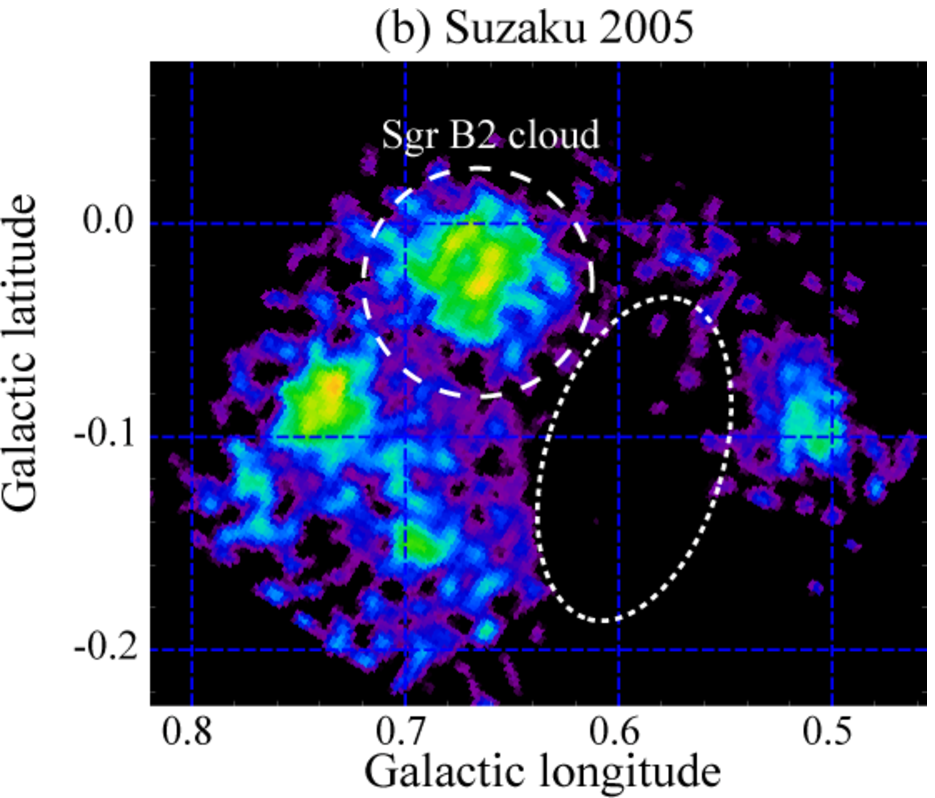}
\end{center}
\caption{6.40 keV maps of  the Sgr~B2 complex taken with Chandra (upper panel) and Suzaku (lower panel).
The coordinates are the Galactic longitude and latitude. Sgr A* is at the right-hand side of these figures.
The data were smoothed with 1-sigma of 10 arcsec. The Sgr~B2 complex
is shown by the solid line in the figure (upper panel), while the Sgr~B2 cloud is given by the dashed line (lower panel).}
\label{fig:6.4 keV-Image}
\end{figure}
As the background, off-plane blank-sky  data were used: north ecliptic pole data for XIS and a distributed blank-sky database for 
ACIS. We made the ARF using real images.
The background-subtracted spectra exhibit three pronounced peaks which represent Fe \emissiontype{I}-K$\alpha$ (6.40~keV), Fe \emissiontype
{XXV}-K$\alpha$ (6.68~keV), and the composite lines of Fe \emissiontype{XXVI}-K$\alpha$ (6.96~keV) and Fe \emissiontype{I}-K$\beta$(7.06~keV). 

Since the Suzaku spectra have better statistics and an accurate line energy determination (Koyama et al. 2007c), 
we derived the best-fit model in the 5--8 keV band with a power-law and four Gaussians, with a fixed energy 
interval  between Fe \emissiontype{I}-K$\alpha$ (6400~eV) 
and K$\beta$ (7058~eV) to the theoretical value (658~eV) \citep{Kaastra1993}. The intrinsic width of the 6.68~keV line was assumed to be 23~eV, 
following Koyama et al. (2007c). This line broadening is due to blending of the resonance, inter-combination and forbidden lines of He-like irons 
(Fe \emissiontype{XXV}-K$\alpha$) and several satellite lines. 
Those of the other lines were assumed to be narrow (fixed to $\sim0$~eV).
The flux ratio of Fe \emissiontype{XXVI}-K$\alpha$ against Fe \emissiontype{XXV}-K$\alpha$ was fixed to be 0.3,
slightly smaller than that of the GCDX (Koyama et al. 2007c).  We estimated this value based on the GCDX data plus an additional contribution 
of the 6.7 keV line from a new SNR candidate, G~0.61+0.01, located in the Sgr~B2 complex region (Koyama et al. 2007b). 
The best-fit spectra and parameters are given in  figure 2 and table 1, respectively. $N_{\rm H}$ is 
$(3.0\pm0.1)\times10^{23}$~H~cm$^{-2}$, determined by the depth of the iron K-edge with the solar iron abundance. 

We also fit the Chandra spectrum, but $N_{\rm H}$ 
 and the flux of the composite lines at $\sim 7$ keV were 
not well determined due to poor statistic above 7 keV energy (see figure 2, upper panel).
We therefore fixed the power-law index and $N_{\rm H}$ to those determined with Suzaku.
We further fixed the line center energies to the Suzaku best-fit values.
Still, the line and edge structures could not  well fitted for the Chandra spectrum, 
indicating the presence of a systematic gain offset. We therefore fine-tuned the energy offset by $-20$ eV, and obtained a nice fit.
The best-fit spectra and parameters are also shown in figure 2 and table 1, respectively. 
The flux ratios between Fe \emissiontype{I}-K$\alpha$ (6.40 keV) and Fe \emissiontype{I}-K$\beta$ (7.06 keV) were  determined to be 0.08
(for the Suzaku data) and 0.13 (for Chandra).

\begin{figure}[!ht]
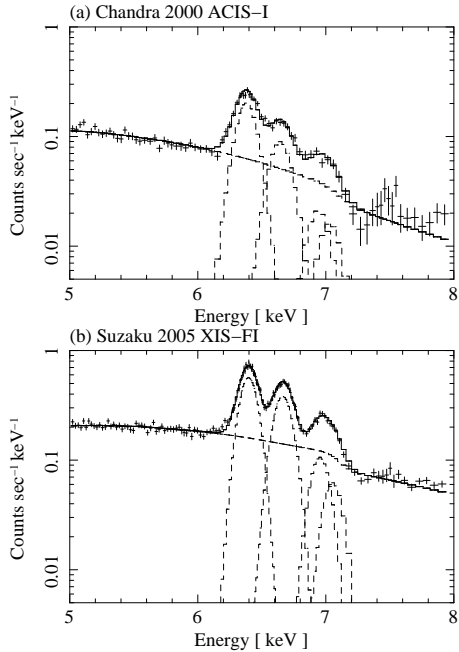
 
\begin{center}
\includegraphics[angle=270,width=60mm]{Fig2a.eps}
\includegraphics[angle=270,width=60mm]{Fig2b.eps}
\caption{X-ray spectra from the Sgr~B2 complex, where the off-plane blank sky spectra were  subtracted. 
Upper panel, Chandra-ACIS;  Lower panel, Suzaku-XIS(FI).
In the spectra, the dashed lines are the best-fit Gaussians for the emission lines; the solid lines are those for the continuum emissions.}
\label{fig:6.4 keV-Spec}
\end{center}
\end{figure}

\begin{table*}[!ht] 
\begin{center}
\caption{Best-fit parameters of the Sgr~B2 complex region.}
\begin{tabular}{llll}
\hline
         & 6.40 keV      & 6.68 keV  &7.06 keV \\
         & $10^{-4}$~photons~cm$^{-2}$~s$^{-1}$  
	 & $10^{-4}$~photons~cm$^{-2}$~s$^{-1}$  
	 & $10^{-4}$~photons~cm$^{-2}$~s$^{-1}$  	 \\
\hline
Suzaku       & $3.2\pm0.2$    & $2.4\pm0.1$ & $0.26^{+0.06}_{-0.03}$ \\
Chandra      & $4.5\pm0.2$    & $2.6\pm0.2$ & $0.60\pm0.16$  \\
\hline
\end{tabular}
\end{center}
\end{table*}

Since the 6.68 keV line is certainly due to the largely extended GCDX
(Koyama et al. 2007c), 
this line flux must be time constant. In fact, the observed flux with Chandra
is ($2.6\pm0.2$)$\times10^{-4}$~photons~cm$^{-2}$~s$^{-1}$, while that of Suzaku is 
($2.4\pm0.1$)$\times10^{-4}$~photons~cm$^{-2}$~s$^{-1}$ ; no difference is found within the statistical error. 
This, in tern, verified the reliability to study any possible flux change
of the adjacent 6.40 keV line between the Chandra and Suzaku observations.
The 6.40 keV flux, in fact, decreased from ($4.5\pm0.2$) $\times10^{-4}$~photons~cm$^{-2}$~s$^{-1}$ in the Chandra (2000) observation
to ($3.2\pm0.2$) $\times10^{-4}$~photons~cm$^{-2}$~s$^{-1}$ in the Suzaku (2005) observations. Compared to the constant
flux of the 6.68 keV line, the variability of the 6.40 keV line is highly confident.
We note that if we relaxed the power-law index for the Chandra spectrum fit, the resultant flux of the  
6.40 keV line became 
 $(4.7\pm0.2)\times10^{-4}$~photons~cm$^{-2}$~s$^{-1}$, while that of the 6.68 keV was 
$(2.6\pm0.2)\times 10^{-4}$~photons~cm$^{-2}$~s$^{-1}$.
These values are the same as those listed in table 1 within statistical errors.

In order to see the 6.40 keV line variability, we depicted the 6.40 
keV photons within the FWHM energy band, and subtracted the underlying continuum flux, which was estimated using the best-fit power-law spectra, shown in figure 2. The results of the 6.40 keV maps are shown in figure 1.  The 
surface brightness is shown by color codes, which were corrected for the exposure time, 
vignetting and detector efficiency for both Chandra and Suzaku. We confirmed the change in the surface brightness of the 6.40 keV line flux between the two observations. 
Since the angular resolution of Suzaku is limited, to compare the morphological change of the Sgr B2 cloud is difficult. We however can
safely state that a drastic change of G~0.570$-$0.018 occurred. 
In the Chandra images, a clear spot near the Sgr B2 cloud is found at the right-hand side of the map (G~0.570$-$0.018: Senda et al. 2002).
In the Suzaku observation, however, G~0.570$-$0.018 has almost disappeared. 

\subsection{The Sgr B2 Cloud}

The brightest spot in figure 1 is the Sgr B2 cloud.
We obtained X-ray spectra of the Sgr B2 cloud from the dashed line circle given in figure 1.
The background was subtracted from the region shown by the dotted line ellipse.  
The background-subtracted spectra of the Sgr B2 cloud are shown in figure 3. The spectra were
simultaneously fitted with a fixed power-law of $\Gamma = 3$ (Koyama et al. 2007b), 
and two Gaussians with a fixed line ratio of Fe\emissiontype{I}-K$\beta$(7.06 keV)/Fe\emissiontype{I}-K$\alpha$(6.40 keV) = 0.1
(the mean value of Suzaku and Chandra for the Sgr B2 complex).  
The best-fit model parameters are given in table 2.  
The mean $N_{\rm H}$ of the Sgr B2 complex is smaller than 
that of the Sgr B2 cloud.
Therefore, the derived 6.40 keV line flux of the Sgr B2 cloud, corrected by the larger $N_{\rm H}$ absorption, becomes larger than that of the Sgr B complex corrected by smaller $N_{\rm H}$ absorption. This, however, does not change the fact that the 6.40 keV line fluxes from  both the Sgr B2 complex and the cloud are time variable.

\begin{figure}[!ht]
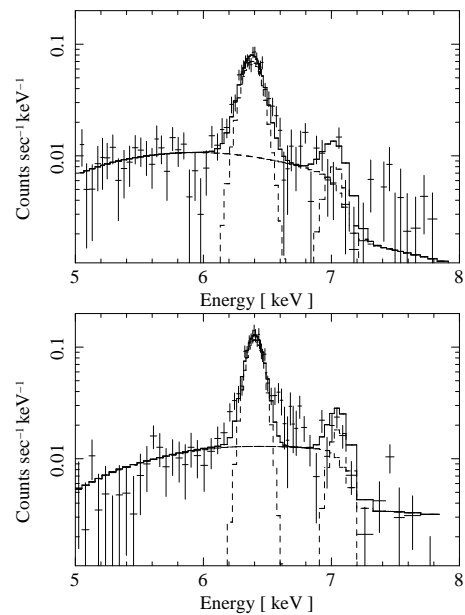
 
\begin{center}
\includegraphics[angle=270, width=60mm]{Fig3a.eps} 
\includegraphics[angle=270, width=60mm]{Fig3b.eps}
\caption{X-ray spectra from the Sgr B2 cloud  taken from the dashed line given in figure 1, 
where the background spectra are taken from the dotted line.
In the spectra, the dashed lines are the best-fit Gaussians for the emission lines, and the solid lines are those for the continuum emissions. The upper panel is Chandra ACIS, while the lower panel is Suzaku XIS(FI).
The excesses at 6.7 keV and 7.0 keV are due to a small contamination of the GCDX.}
\label{fig:6.4 keV-Spec}
\end{center}
\end{figure}

\begin{table*}[!ht]
\begin{center}
\caption{Best-fit parameters of the Sgr B2 cloud.}
\begin{tabular}{llll}
\hline
         & 6.40 keV  & Equivalent width ($EW$)   &  $N_{\rm H}$ \\
         & $10^{-4}$~photons~cm$^{-2}$~s$^{-1}$  
	 & keV
	 & $10^{23}~{\rm H~cm}^{-2}$\\
\hline
Suzaku& $2.8^{+0.1}_{-0.5}$ & $1.6^{+0.1}_{-0.3}$  &  $11\pm1$ \\
Chandra& $5.5^{+0.4}_{-0.3}$ & $1.6^{+0.2}_{-0.2}$   & fixed to Suzaku\\
\hline
\end{tabular}
\end{center}
\end{table*}

\section{Discussion}

We detected a strong K$\alpha$ line at 6.40 keV, and a weak K$\beta$ line at 7.06 keV
from both the Sgr B2 complex and the Sgr B2 cloud. 
These lines are produced when the K-shell electrons in iron atoms become vacant; subsequently a
L-shell/M-shell electron decays to this vacant level. 
The flux ratio (K$\beta$/K$\alpha$) is consistent with 0.1.  This ratio indicates that the M-shell is fully occupied by electrons,
and hence the iron atom is nearly neutral \citep{Kaastra1993}.
Then, a question is how to make the
K-shell electron vacant (an inner-shell ionization).  The inner-shell ionization can be
either by X-rays or by energetic electrons.

We derived  Fe K$\alpha$ $EW$ of ~1.6 keV from the Sgr B2 cloud.
This can be produced via the
irradiation of X-rays, whereas collisional excitation by electrons could 
produce an $EW$ of only 300--500 eV for the solar abundances (Tatischeff 2003).  
One may argue, however, that because the GC typically 
has abundances of 1.5--3 times solar, the observed $EW$ might 
be consistent with electron origin, if the metallicity of Sgr B2 is ~3-times solar. 
This possibility can not be entirely excluded.

We also observed a large $N_{\rm Fe}$ value (equivalent $N_{\rm H}$ of $\sim10^{24}~{\rm H~cm}^{-2}$ for the Sgr B2 cloud, assuming the solar abundance). Since this value is one order of
magnitude larger than the general absorptions to the GC sources
($\sim6\times10^{22}~{\rm H~cm}^{-2}$ ) \citep{Sakano2002}, most of the absorption is due to local origin, the Sgr B2 cloud.
Electrons are very difficult to explain this large $N_{\rm H}$,
because, at the energy of the maximum cross section of inner-shell  ionization
(10--100 keV), they can go into the molecular cloud only in the order of $N_{\rm H}$ of $10^{21-22}~{\rm H~cm}^{-2}$
\citep{Tatischeff2003}.
One argument to support the electron origin is that electron bombarding takes place at the rear side of the cloud. Although it seems artificial, this argument can not be entirely excluded.

We found  more direct evidence for the X-ray irradiation origin: the time
variability of the 6.40 keV line flux.
The linear size across the Sgr B2 complex is about 20 light-years, but the bright part (the Sgr B2 cloud) is limited  to
within 10 light-years. 
The 6.40 keV flux changed by factor 2 within 5 years. This time scale is comparable to the light-crossing time
of the cloud. Electrons having  energies at the maximum cross section of the Fe \emissiontype{I} inner-shell ionization 
(10--100 keV) can not move as fast as the speed of light.  Charged particles other than 
electrons are even more difficult as the origin.

A unique solution to explain the spectrum and time variability of the flux and the morphology is: the Sgr 
B2 cloud absorbs variable X-rays above the 7.1 keV edge energies, and simultaneously re-emits the fluorescent 
6.40 keV lines.  The 6.40 keV line flux can be estimated from the solid angle of the Sgr B2 cloud, 
the fluorescent yield of neutral iron (0.34) and the depth of the K-shell edge above 7.1 keV (Murakami et al. 2000, 2001).  
As a result, the observed 6.40 keV flux surely requires a near-by bright X-ray source.

The second question is then: where is the bright X-ray source ?  
From the flux change and its time scale, this putative transient source must be bright for more than 5 years, but show a  flux decrease  by a factor 2. It should be brighter than 10$^{37}$ ergs~s$^{-1}$, if the source is
located at a comparable distance of 10--20 pc as the Sgr B2 size.  
The GC region has been surveyed by many satellites, but no X-ray source brighter than 
10$^{37}$ergs~s$^{-1}$ (5 years average) has been found.  
Near the GC, neutron-star or black-hole binary sources occasionally 
flare-up to more than 10$^{37}$ergs~s$^{-1}$, but the flare durations are typically only a few months.  

Thus, the  most-probable source to exhibit bright and relatively long-lived 
X-rays is a massive black hole, Sgr A$^*$. The required luminosity to account for the Chandra flux of the Sgr B2 cloud is 
estimated to be $2\times 10^{39}$ ergs s$^{-1}$ (Murakami et al. 2001), while that for the Suzaku  is about half.  
This idea has been proposed based on the early 
ASCA and Chandra observations.  New evidence of the large-scale ($\sim$10--20 light-years), long-term ($\sim$ 5 years) flux 
variation of the 6.40 keV line gives decisive support for this early idea. 
Our GC black-hole Sgr A$^*$ shows frequent short-term ($\sim$1 hour) and low-level ($\sim$10--100 times of the quiescent 
level) flares \citep{Baganoff2001}. 
More than a few hundred years ago, Sgr A$^*$  had been very active in X-rays; at $\sim$300 
years ago it was 10$^6$ times brighter than the present value, and decayed to less than half after $\sim$5 years.  
The X-rays hit the Sgr B2 cloud after $\sim$300 years of travel.  
The cloud re-emitted the 6.40 keV photons, like a time delayed-echo. The echo is now just arriving at  Earth, while  Sgr A$^*$ 
is falling into  a quiescent state.

\bigskip
The authors thank all of the Suzaku team members, especially Y. Hyodo, H. Uchiyama, H. Nakajima, H. Yamaguchi, 
and H. Mori for their comments, support and useful information on the XIS performance.
T.I. is supported by JSPS Research Fellowship for Young Scientists.
This work is supported by a Grant-in-Aid for the 21st Century COE "Center for Diversity and Universality in Physics" from the Ministry of Education, Culture, Sports, Science and Technology (MEXT) of Japan.

\end{document}